\documentclass[aps,twocolumn,nofootinbib,preprintnumbers,groupedaddress,letterpaper]{revtex4}

\usepackage[dvips]{color}
\usepackage[normalem]{ulem}
\usepackage{amsmath}
\usepackage{enumerate}
\usepackage{amsfonts}
\usepackage{epsfig}
\usepackage{yfonts}

\usepackage{feynmp}
\usepackage{graphicx}
\usepackage{graphicx}
\usepackage{bm}
\usepackage{subfigure}

\newcommand\comment[1]{}

\newcommand\reissner{Reissner-Nordstr\" om }
\newcommand\stuck{St\" uckelberg }

\newcommand\ep{\epsilon}

\newcommand\om{\omega}

\newcommand\ov{\over }

\def\le{\left}
\def\ri{\right}

\def\({\left(}
\def\){\right)}
\def\[{\left[}
\def\]{\right]}
\def\<{\langle}
\def\>{\rangle}

\def\Tr{\mathop{\rm Tr}}

\newcommand\half{{\ensuremath{\frac{1}{2}}}}
\newcommand\p{\ensuremath{\partial}}

\newcommand\field[1]{{\ensuremath{\mathbb{{#1}}}}}

\newcommand{\RR}{\field{R}}

\newcommand{\be}{\begin{equation}}
\newcommand{\ee}{\end{equation}}
\newcommand{\bea}{\begin{eqnarray}}
\newcommand{\eea}{\end{eqnarray}}
\newcommand{\bwt}{\begin{widetext}}
\newcommand{\ewt}{\end{widetext}}

\newcommand{\bi}{\begin{itemize}}
\newcommand{\ei}{\end{itemize}}
\newcommand{\ben}{\begin{enumerate}}
\newcommand{\een}{\end{enumerate}}
\newcommand{\bca}{\begin{cases}}
\newcommand{\eca}{\end{cases}}
\newcommand{\bln}{\begin{align}}
\newcommand{\eln}{\end{align}}
\newcommand{\bst}{\begin{split}}
\newcommand{\est}{\end{split}}

\begin{document}

\preprint{CERN-PH-TH/2013-357}

\title{Holography without translational symmetry }

\author{David Vegh}
\email{david.vegh@cern.ch}
\affiliation{\it Theory Group, Physics Department, CERN, CH-1211 Geneva 23, Switzerland}

\date{\today}
\begin{abstract}

We propose massive gravity as a holographic framework for describing a class of strongly interacting quantum field theories with broken translational symmetry.
Bulk gravitons are assumed to have a Lorentz-breaking mass term as a substitute for spatial inhomogeneities. This breaks momentum-conservation in the boundary field theory.
At finite chemical potential, the gravity duals are charged black holes in asymptotically anti-de Sitter spacetime.
The conductivity in these systems generally exhibits a Drude peak that approaches a delta function in the massless gravity limit. Furthermore, the optical conductivity shows an emergent scaling law: $|\sigma(\om)| \approx {A \ov \om^{\alpha}} + B$. This result is consistent with that found earlier by Horowitz, Santos, and Tong who introduced an explicit inhomogeneous lattice into the system.

\end{abstract}

\maketitle

\section{Introduction}

In systems with perfect translational symmetry, the particles cannot dissipate their momentum. Consequently, in the presence of a finite density of charge carriers, there is a delta function in the AC conductivity at zero frequency. The DC conductivity is therefore infinitely large. This unwanted result can be avoided by treating the charge carriers in the probe limit (i.e. as a small part in a larger system of neutral fields where they can dump momentum), or by introducing spatial inhomogeneities thereby breaking translational invariance explicitly.

Once momentum dissipation has been introduced into the system, the results will be finite. This can be demonstrated by the Drude model of conductivity: a phenomenological theory that treats the charge carriers as classical particles which can bounce off a background ion lattice. The equation of motion,
\be
  {d \over dt} \vec p(t) = e \vec E - { \vec p(t) \ov \tau}
\ee
where $\tau$ is the mean free time between collisions, $q$ is the electron's charge, $\vec E$ is the background electric field driving the current. The DC conductivity is then finite
\be
  \vec j = {ne^2 \tau \ov m^*} \vec E = \sigma_{DC} \vec E
\ee
where $n$ is the number density and $m^*$ the effective electron mass. In order to compute the AC conductivity, one turns on a time-dependent electric field with angular frequency $\omega$. This yields
\be
 \sigma(\omega) = {\sigma_{DC} \over 1-i\omega \tau}
\ee
The pole is shifted to the lower half plane and the DC conductivity is finite.

The transport properties of a large class of metals are reasonably well described by the Drude model. There are materials, however, whose optical conductivity deviates from the simple Drude formula.
In one class of high temperature superconductors, for instance, the observed conductivity in the normal phase follows a power law $|\sigma(\om)| \propto (-i\om)^{-2/3}$ (see \cite{PhysRevB.49.9846, 2003Natur.425..271M}). These systems are strongly coupled and there is no simple quasiparticle-based Fermi liquid description.

In the past fifteen years, there has been much progress in understanding certain strongly interacting quantum field theories in the 't Hooft limit using the AdS/CFT correspondence \cite{Maldacena:1997re, Gubser:1998bc, Witten:1998qj}.
Translational symmetry breaking has been studied and Drude(-like) peaks were discovered \cite{2007PhRvB..76n4502H, Hartnoll:2007ip, Hartnoll:2008hs, Hartnoll:2009ns, Hartnoll:2012rj}.
Recent numerical calculations in holographic lattice systems gave evidence for universal non-Drude frequency scalings \cite{Horowitz:2012ky, Horowitz:2012gs}.

Holography in itself is not doing any coarse-graining, therefore such calculations on inhomogeneous backgrounds  require the solution of partial differential equations\footnote{For conductivity calculations on a {\it homogeneous} (Bianchi VII) space, see \cite{Donos:2012js}. For other related works, see \cite{Faulkner:2010zz, Liu:2012tr, Karch:2007pd}.}.  This motivates the main goal of this paper which is to build a {\it framework for translational symmetry breaking and momentum dissipation in holography} without the need for complex numerical computations.

\section{Holographic matter}
\label{holo:section}

In order to describe the holographic dual of strongly coupled matter, we are going to use a set of minimal ingredients, namely, Einstein-Hilbert action with a gauge field and a (negative) cosmological constant
\be
 S = {1 \ov 2 \kappa^2} \int d^{4} x \, \sqrt{-g} \le[R + {6 \ov L^2} - {L^2 \ov 4} F_{\mu\nu} F^{\mu\nu} \ri]
\ee
The equations of motion are solved by the following $AdS$-\reissner geometry~\cite{Romans:1991nq, Chamblin:1999tk}
\be
  ds^2 = L^2 \le( {dr^2 \ov f(r) r^2} + {-f(r)dt^2 + dx^2 + dy^2 \ov r^2}  \ri)
\ee
\be
   A(r) = \mu\le(1 - {r\ov r_h}\ri)dt
\ee
where the emblackening factor is
\be
  f(r) = 1  - M r^3+Q^2 r^4     
\ee
This is a charged black brane with a horizon at $r_h$ which is the smallest positive root of $f(r)$. Using the AdS/CFT dictionary, $\mu $ is identified with the chemical potential of the boundary theory.
In the zero-temperature limit, the near-horizon metric becomes $AdS_2 \times \RR^{2}$ which governs much of the low-energy physics.

This geometry describes a translationally invariant state. The electrical conductivity in the boundary theory can be computed by means of the Kubo formula
\be
  \sigma(\om) = {1\ov i \om} \langle J_x(\om) J_x(-\om) \rangle
\ee
where $\langle\ldots\rangle$ denotes the retarded Green's function. The currents are dual to the bulk gauge field. Thus, we need to compute the boundary-to-boundary two-point function of the gauge field $a_x$ at finite frequency and zero spatial momentum.
At non-zero charge density, due to the background gauge field profile, $a_x(r)$ mixes with the graviton $g_{tx}(r)$. Let us rescale the variables such that $r_h=1$. To first order in the perturbations, the Maxwell equation is
\be
  2 \omega^2 a_x+f \left[-\mu (\tilde g_t^x)'+2 \left(a'_x f'+f a''_x\right)\right] = 0
\ee
where $\tilde g_t^x \equiv g_{tx} g^{xx}$, and at the linear level $\tilde g_t^x \approx g_t^x$.
The $r-x$ component of Einstein's equations
\be
4 \mu r a_x -{  (\tilde g_t^x)' \ov r}=0
\ee
The $t-x$ component of Einstein's equations follows from the $r-x$ equation above (by taking the derivative). We impose ingoing boundary conditions at the horizon. This corresponds to computing retarded Green's functions \cite{Son:2002sd}.
The two coupled differential equations can then be solved (e.g. by the method of matched asymptotic expansions or numerically). In the ultraviolet region, $a_x \sim a_+ + a_- r + \mathcal{O}(r^2)$, and the Green's function is given by the ratio $G = {a_- \ov a_+}$.

The result contains a Dirac-delta function in the real part of the conductivity. This is a consequence of the Ward identities for the translational symmetry.
In the following, we resolve this delta function by letting momentum dissipate.

For more information on holographic matter, see the review papers \cite{Hartnoll:2009sz, McGreevy:2009xe, Herzog:2009xv}.

\section{Breaking translational invariance}

In the charged black brane background from the previous section, Ward identities for translational invariance in the $x$ direction imply a shift symmetry in the $g_t^x$ field. This is why only derivatives of $\tilde g_t^x$ arose in the equations of motion. In order to to break translational symmetry, the shift symmetry must be broken. The simplest option is to add a mass term for the graviton,
\be
\mathcal{L}_I = \sqrt{-g} \, m^2 (\delta g_{tx})(\delta g^{tx})
\ee
where indices are raised using the diagonal background metric. Since the background is diagonal, $\delta g_{tx} = g_{tx}$.
The graviton mass term produces a linear $g_{tx}$ term in the $t-x$ component of Einstein's equations. However, the $r-x$ component is unchanged and the two equations are now incompatible unless the $g_{rx}$ graviton component is also non-vanishing. This component carries an extra degree of freedom.\footnote{An interpretation of this field is the following. In a semi-holographic effective theory \cite{Faulkner:2010tq, Nickel:2010pr, Faulkner:2009wj}, one separates the bulk spacetime into a UV and an IR region,
\be
  S =  S_\textrm{UV}(G_{IJ}, g_{IJ}) + S_\textrm{IR}(g_{IJ})
\ee
where $G_{IJ}$ is the UV boundary value of the metric, and $g_{IJ}$ is its value at a fixed  intermediate cutoff scale, and $I,J \in \{t,x,y\}$. The action is invariant under coordinate transformations that change either $g$ or $G$.
In the low energy theory the two symmetry groups are broken down to the diagonal. Finally, the Goldstone bosons corresponding to the broken axial symmetry are the radial integrals of the $g^{ri}$ fields in the UV region.
}

\subsection{Non-linear massive gravity}

If we add generic mass terms for the gravitons on a given background, then the theory will be plagued by various instabilities, sometimes at the non-linear level. Recently, the authors of \cite{deRham:2010kj} constructed a theory where the Boulware-Deser ghost \cite{1972PhRvD...6.3368B} was eliminated by introducing higher order interaction terms into the Lagrangian. (See also earlier works \cite{ArkaniHamed:2002sp, deRham:2010ik, Porrati:2001db, Apolo:2012gg, tHooft:2007bf}. For a recent review of massive gravity in this context, see \cite{Hinterbichler:2011tt}.)

In 3+1 dimensions, the theory has two dimensionless parameters, and it also depends on a fixed rank-2 symmetric tensor $f$, the {\it reference metric}. The usual dynamical metric will be denoted by $g$. For our purposes, we include a cosmological constant and a Maxwell field,
\be
  \label{eqn:nonlinaction}
  S = {-1 \ov 2 \kappa^2} \int d^{4} x \, \sqrt{-g} \le[R  + \Lambda -{L^2\ov 4} F^2 + m^2 \sum_{i=1}^4 c_i \, \mathcal{U}_i(g,f) \ri]
\ee
where $c_i$ are constants, $\mathcal{U}_i$ are symmetric polynomials of the eigenvalues of the $4\times4$ matrix ${\mathcal{K}^\mu}_\nu \equiv  \sqrt{g^{\mu\alpha}f_{\alpha\nu}}$
\bea
    & \mathcal{U}_1  = & [\mathcal{K}] \\
    \nonumber
    & \mathcal{U}_2  = & [\mathcal{K}]^2 - [\mathcal{K}^2 ] \\
    \nonumber
    & \mathcal{U}_3  = & [\mathcal{K}]^3 - 3[\mathcal{K}] [\mathcal{K}^2]+2[\mathcal{K}^3]  \\
    \nonumber
    & \mathcal{U}_4  = & [\mathcal{K}]^4 - 6[\mathcal{K}^2] [\mathcal{K}]^2+8[\mathcal{K}^3] [\mathcal{K}]+3[\mathcal{K}^2]^2 -6[\mathcal{K}^4]
\eea
The square root in $\mathcal{K}$ is understood to denote matrix square root, i.e. ${(\sqrt{A})^\mu}_\nu{(\sqrt{A})^\nu}_\kappa = {A^\mu}_\kappa$. Rectangular brackets denote traces: $[\mathcal{K}] \equiv {\mathcal{K}^\mu}_\mu$.
As $m\to 0$, we recover the translational invariant action in section \ref{holo:section}.

If the reference metric is flat, we can express it via a coordinate transformation $\phi^a(x)$ using $\eta_{ab}$,
\be
  f_{\mu\nu} = \p_\mu \phi^a \p_\nu \phi^b \eta_{ab}
\ee
Different choices for the $\phi^a$ \stuck fields correspond to different gauges. The { unitary} (or physical) gauge is defined simply by $  \phi^a = x^\mu \delta^a_\mu$.

In this paper we will be interested in the case of a {\it spatial reference metric} (in the basis $(t,r,x,y)$)
\be
  f_{\mu\nu} =  (f_{sp})_{\mu\nu} = \textrm{diag}(0,0,1,1)
\ee
Note that the action remains finite since it only contains non-negative powers of $f_{\mu\nu}$.
The reason for using this singular metric becomes clear if we perform a coordinate transformation $\phi^a(x)$  which preserves the $x$ and $y$ coordinates,
\be
  \phi^{t,r} = \phi^{t,r}(t,r) \qquad \phi^x = x \qquad  \phi^y = y
\ee
The reference metric and the action stay the same. This means that the  spatial graviton mass term $m^2\mathcal{U}(g, f_{sp})$ preserves general covariance in the $t$ and $r$ coordinates, but breaks it in the two spatial dimensions. This is exactly what we need.

Since the reference metric is spatial, we only need two \stuck fields $\phi^x$ and $\phi^y$. They can be thought of as maps used in the Lagrangian representation of the degrees of freedom in a solid. In analogy to crystals, these dofs may be called {\it ions}.
Perturbations of the \stuck fields are the {\it phonons}
\be
  (\phi^x, \phi^y) = \le(x + \pi^x, \, y + \pi^y  \ri)
\ee
In this interpretation, the bulk is filled with a homogeneous solid  \cite{Dubovsky:2005xd}.
Due to a gauge symmetry we can either set $g_{ri}=0$ and have $\pi^i \ne 0$ or vice-versa.

In the ADM formulation \cite{Arnowitt:1962hi}, the metric is parametrized in the following way: $ N = (-g^{00})^{-1/2}$, $ N_i = g_{0i} $, $ \gamma_{ij} = g_{ij}$. Furthermore, let us define  $\gamma^{ij} \gamma_{jk} = \delta^i_k$ and $N^i = \gamma^{ij} N_j$. In terms of these variables, the spatial graviton mass term assumes the explicit form\footnote{There are only two terms, since the spatial gauge only allows for two non-zero eigenvalues for the matrix $\mathcal{K}$.}
\be
  \label{eqn:nonlinexplicit}
  m^2 \mathcal{U}_{sp.} = m^2(\alpha \mathcal{V}_1+ \beta \mathcal{V}_2)  
\ee
{
\bea
    \nonumber
    & \mathcal{V}_1  =  \sqrt{\Tr(\tilde\gamma^{-1}\tilde f)- \tilde f_{ij} { N^i N^j \ov N^2}+ \mathcal{V}_2} & \\
    \nonumber
    & \mathcal{V}_2  =   \sqrt{\det(\tilde\gamma^{-1}\tilde f)} \, \sqrt{1 -  \tilde\gamma_{ij} { N^i N^j \ov N^2 } } &
\eea
}
where $\tilde\gamma^{-1} = \binom{\gamma^{xx} \ \gamma^{xy}}{\gamma^{xy} \ \gamma^{yy}}$, $\tilde\gamma\equiv(\tilde\gamma^{-1})^{-1}$, and  the spatial submatrix of the reference metric was kept in a general form: $\tilde f = \binom{f_{xx} \ f_{xy}}{f_{xy} \ f_{yy}}$. $\alpha$ and $\beta$ are two parameters (equal to $c_1$ and $c_2$ in eqn. (\ref{eqn:nonlinaction})).

Both $\mathcal{V}_{1,2}$ are invariant under spatial rotations. Note that they do not contain $N^r$ and thus the corresponding constraint in the ADM formulation is preserved.
Note that the $g^{ri}$ components do not show up in these terms either\footnote{\ldots even though $N^i$ do appear. This is due to the asymmetric parametrization of the metric w.r.t. the $r$ and $t$ coordinates.}.

Instead of using $\mathcal{K}$, we could consider mass terms made out of ${\mathcal{\tilde K}_\mu}{}^\nu \equiv  \sqrt{g_{\mu\alpha} f^{\alpha\nu}}$ where the inverse reference metric is set to $ f^{\mu\nu} = \textrm{diag}(0,0,1,1)$.
It is easy to check that the new functions $\mathcal{U}_i(\mathcal{\tilde K})$ do not contain the $N^\mu$ lapse and shift fields. However, unlike  $\mathcal{U}_i(\mathcal{K})$, they are functions of $\gamma^{r\mu}$. We will not consider this option here.


In this paper, we will not consider the delicate question of ghosts and tachyons. These investigations typically depend on the background metric and the choice of parameters (see e.g. \cite{Deser:2001wx}).
We just note here that the Hamiltonian constraint can be preserved by redefining the fields $N^i \to n^i$ using a transformation that is linear in $N$
\be
  N^i = n^i +  d^i(n^i, \tilde\gamma) N
\ee
and then choosing $d^i$ such that $\sqrt{-g}\,  \mathcal{U} \propto N \mathcal{U}(n^i, N, \tilde\gamma) $ is linear in $N$. For instance, assuming $\tilde f = \textrm{Id}_{2\times 2}$, for $\mathcal{V}_2$ this can be done by setting $N^x = n^x$ and $N^y = n^y +  (n^x)^{-1} \det(\tilde\gamma)^{-1/2}\sqrt{\tilde\gamma_{ij} n^i n^j }N$. Then, the mass term becomes linear in $N$
\be
  N \mathcal{V}_2 = \det(\tilde\gamma)^{-1/2}\sqrt{\tilde\gamma_{ij} n^i n^j } + \le(\tilde\gamma^{xy} - \tilde\gamma^{xx}{n^y \ov n^x} \ri) N
\ee
Thus, the Boulware-Deser ghost is eliminated from the theory. For more on this, see \cite{Hassan:2011vm, Hassan:2011hr, Hassan:2011tf}.

\section{Gravity background}
\label{sec:bg}

In the following, we study the massive gravity action (\ref{eqn:nonlinaction}) with the explicit mass term (\ref{eqn:nonlinexplicit}) and set
\be
  f_{\mu\nu} =  \textrm{diag}(0,0,F^2,F^2)
\ee
We will be looking for charged black brane solutions.

We obtain the following solutions to the equations of motion\footnote{Note that using the reference metric, a new invariant can be defined: $I^{ab} = g^{\mu\nu}\p_\mu \phi^a \p_\nu \phi^b$. In unitary gauge,  $I^{ab} = g^{\mu\nu}\delta^a_\mu\delta^b_\nu$, which is singular if the inverse metric is divergent. This presumably leads to perturbative instabilities. Hence, in order to describe a black hole, one typically needs an ansatz for the geometry in which the metric has no horizon singularities. On the other hand, in our spatial gauge  $I^{ab}$ does not depend on $g^{tt}$ since $f_{t\mu} = 0$. Thus, we will be able to use simple coordinate systems.}
\be
  ds^2 = L^2 \le( {dr^2 \ov f(r) r^2} + {-f(r)dt^2 + dx^2 + dy^2 \ov r^2}  \ri)
\ee
\be
   A(r) = \mu\le(1 - {r\ov r_h}\ri)dt
\ee
where the emblackening factor is
\be
  f(r) = 1 +\alpha F \frac{L m^2}{2}     r  + \beta F^2 m^2  r^2  - M r^3+\frac{\mu^2}{4 r_h^2 } r^4
\ee
For the equations of motion, see Appendix A.
The horizon is located at $r_h$ where both $f(r)$ and $A(r)$ vanish.
\comment{We can rescale the geometry by taking
\bea
\nonumber
&  r \to  r_h r \qquad M \to -{1+Q^2 \ov r_h^3} \qquad \mu \to {\sqrt{2} Q L / r_h^2} & \\
&  \om \to \om L / r_h \qquad g_{rx} \to {L\ov r_h} g_{rx} &
\eea
which puts the horizon to $r=1$.
}

There are two dimensionless parameters: $\alpha$ and $\beta$. $F$ and $m$ are redundant parameters and are only included for convenience.
Note that if $m=0$ or $F=0$, then the solution reduces to the $AdS$-\reissner solution of massless Einstein-Maxwell theory.
The temperature is given by
\be
T = { 1 \ov 4 \pi r_h} \le(3 - \le({ \mu \, r_h \ov  2}\ri)^2 + F r_h m^2(\alpha L+ \beta F r_h) \ri)
\ee
Note that whenever $\mu < 2 m F \sqrt{\beta}$, the function $T(r_h)$ has a minimum at
\be
  r_\textrm{min} = { \sqrt{3} \ov \sqrt{ m^2 F^2 \beta-{\mu^2\ov 4}}}
\ee
and there is a corresponding minimal temperature $T=T(r_\textrm{min})$. Below the critical size the black brane is unstable.

\begin{figure}
\begin{center}
\includegraphics[scale=0.6]{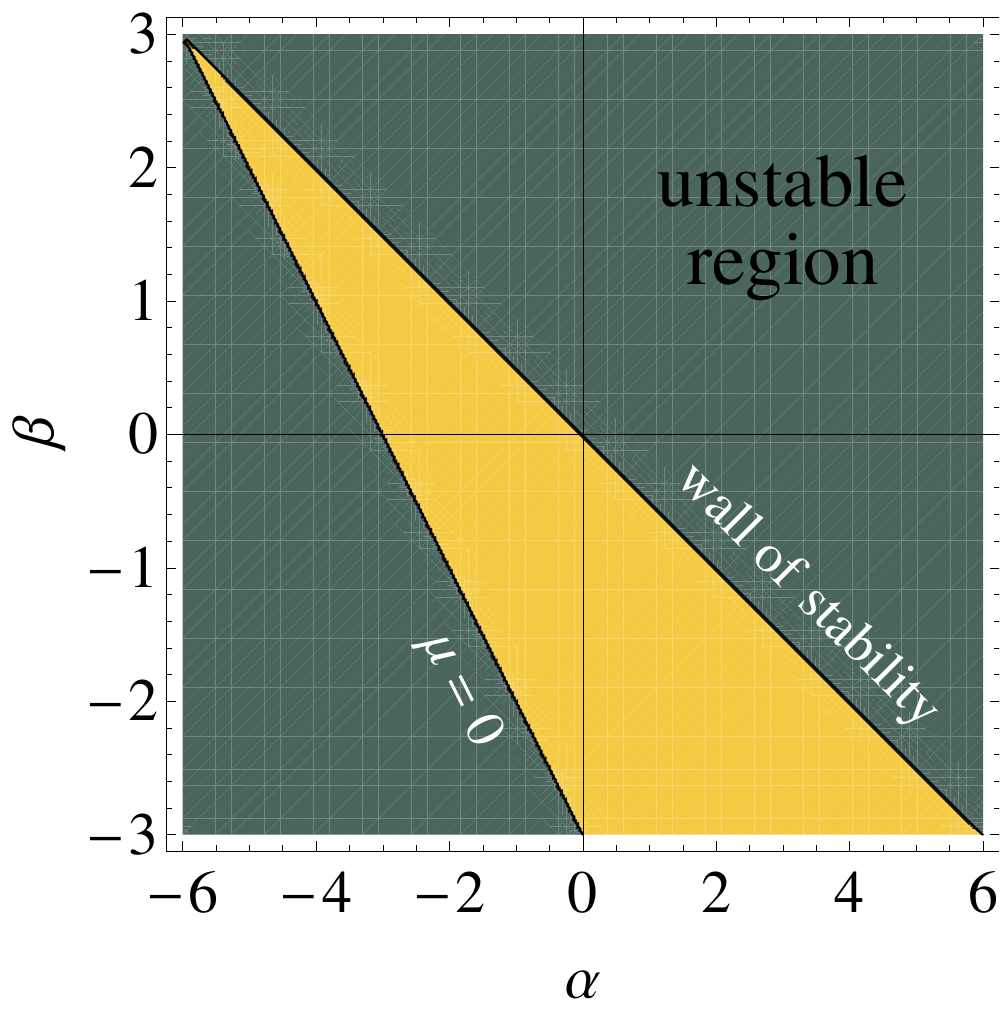}
\caption{\label{fig:psc} Stability in parameter space. We set $r_h = L = m^2 = F = 1$ for the plot. Above the ``wall of stability'' $\beta = - {L \alpha \ov 2 F r_h}$ the entropy density is larger than the usual value (``$S=A/4$'') and numerical results indicate an instability. On the line $\beta = -{3 + F L m^2 r_h \alpha \ov F^2 m^2 r_h^2}$, the maximal value of the chemical potential is zero. Between these two lines (yellow region) the system can be stable. The lines cross at $(\alpha,\beta)=(-{6\ov F L m^2 r_h},{3\ov F^2 m^2 r_h^2})$. Beyond this point there may still be stable points.
}
\end{center}
\end{figure}

The geometry describes a finite density state with the entropy, charge and energy densities respectively given by
\bea
 \nonumber
 &
 s = \frac{4 L^2 \pi}{r_h^2 \kappa^2}\cdot\frac{1+\alpha\frac{2 F L m^2 r_h }{12+r_h^2 \mu ^2}}{1-\beta\frac{4 F^2 m^2 r_h^2  }{12+r_h^2 \mu ^2}} &\\
 \nonumber
 &
 \rho =   { L^2 \mu   \ov \kappa^2 r_h}\le( 1+\frac{F m^2 r_h (L \alpha +2 F r_h \beta )}{12+r_h^2 \left(\mu ^2-4 F^2 m^2 \beta \right)} \ri)  &\\
 \nonumber
 &\ep = \frac{L^2}{4 r_h^2 \kappa ^2} \left(8 M r_h^2+F L m^2 \alpha +\frac{4 F^2 m^4 r_h \left(L \alpha +2 F r_h \beta \right)^2}{2 M r_h^3- F m^2 r_h \left(L \alpha +4 F r_h \beta \right)+4 }\right)&
\eea
These quantities were obtained by computing the action for the (Euclidean) background with the UV divergences removed. This defines the grand canonical ensemble from which we get the above results (see \cite{Chamblin:1999tk} for similar calculations on $AdS$-\reissner backgrounds).
By construction, $\ep, s$ and $\rho$ satisfy the first law of thermodynamics
\be
 d \ep = T ds + \mu d \rho \ .
\ee
Interestingly, the entropy density differs from its usual value $s_0 \equiv \frac{4 L^2 \pi}{r_h^2 \kappa^2}$, unless
\be
  \beta = -{L \alpha\ov 2 F r_h}
\ee
In this case, the energy and charge densities also simplify. This line will be called the {\it wall of stability} for reasons that will become clear later.

Let us fix $r_h = 1$. At fixed graviton masses, $\mu$ is maximized if we set $T=0$. On the line
\be
\beta = -{3 + F L m^2 r_h \alpha \ov F^2 m^2 r_h^2}
\ee
the maximal value of the chemical potential is zero (here $\rho = 0$ as well). This will be called the ``$\mu=0$'' line.

The ground state entropy is found to be
\be
 s(T=0) =  \frac{4 L^2 \pi}{r_h^2 \kappa^2}\le( 1+\frac{m^2 \left(F L r_h \alpha +2 F^2 r_h^2 \beta \right)}{2 L \left(F m^2 r_h \alpha + 6/L \right)} \ri)
\ee
On the ``$\mu=0$'' line it is equal to $s_0 / 2$. It would be interesting to find an interpretation of these results.

\section{Conductivity}

In order to compute the conductivity, we perturb the background\footnote{
Equivalently, one can also consider the perturbation
\be
  ds^2 \to ds^2 + g_{tx}(r)e^{i\omega t} \qquad \pi^x = \pi^{x}(r)e^{i\omega t}
\ee
\be
  A(r) \to A(r) + a_x(r)e^{i\omega t}dx
\ee}
\be
  ds^2 \to ds^2 + g_{tx}(r)e^{i\omega t} + g_{rx}(r)e^{i\omega t}
\ee
\be
  A(r) \to A(r) + a_x(r)e^{i\omega t}dx
\ee
The equations of motion are presented in Appendix B.
The graviton mass does not appear in the Maxwell equation. Its effects are communicated to the gauge field only through the coupling to the graviton fields.

From the equations we can read off the masses of the $g_{rx}$ and $g_{tx}$ fields. They are equal and depend on the radial direction
\be
  \nonumber
  \overline m^2(r) = {r F \ov 2 L^2} \le(  \alpha L  + \beta  F r  \ri)m^2
\ee
Using this formula the temperature can be rewritten as
\be
T = { 1 \ov 4 \pi r_h} \le(3 - \le({ \mu \, r_h \ov  2}\ri)^2 + 2L^2 \overline m^2(r_h) \ri)
\ee

At $T=0$, the scaling dimension of $a_x$ in the infrared $AdS_2$ is given by
\be
  \Delta = \frac{1}{2}+\frac{1}{2} \sqrt{17-\frac{16 \left(6+L m^2 r_h \alpha \right)}{12+m^2 r_h (3 L \alpha +2 r_h \beta )}}
\ee
From this we get $\Delta = 2$ (the $m=0$ result) only on the wall of stability where $\beta = -{L \alpha\ov 2 F r_h}$. On the ``$\mu=0$'' line the formula gives  $\Delta = 1$.

We can eliminate $g_{tx}$ from the equations and obtain two coupled second order equations for $g_{rx}$ and $a_x$. These two equations have been used for the numerical calculations in this paper. At the horizon, we impose infalling boundary conditions
\be
  a_x(r), \ g_{rx}(r) \propto (r_h-r)^{-{ i \omega \ov 4 \pi T }}
\ee
We set normalizable UV boundary conditions for the $g_{rx}$ field\footnote{If we intend to use phonon fields instead of the $g_{rx}$ graviton, then we may set an (equivalent) Dirichlet boundary condition on $\pi'(r)$ at the UV boundary. (Due to a shift symmetry, $\pi(r)$ itself does not appear in the equations of motion.) }.
This determines the wavefunctions up to a constant factor.
We proceed to read off the Green's function: $a_x \sim a_+ + a_- r + \mathcal{O}(r^2)$ near the boundary, and then $G(\om) = {a_- \ov a_+}$ as earlier. Finally, the Kubo formula gives the conductivity: $\sigma(\om) = G(\om)/(i\om)$.

\begin{figure}
\begin{center}
\includegraphics[width=80mm]{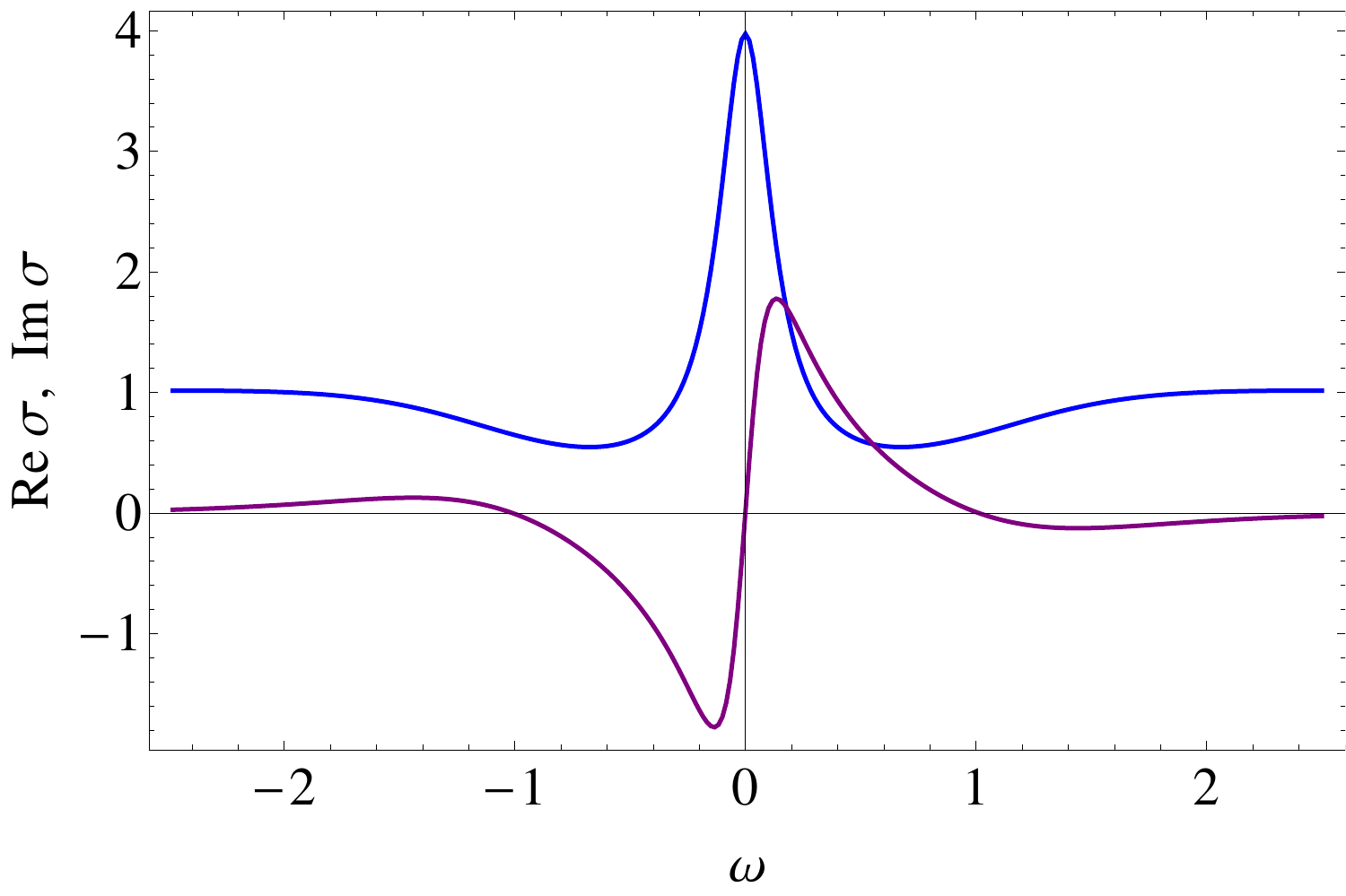}
\caption{\label{fig:pic1} Drude peak in the conductivity. The real and imaginary parts are drawn in blue and purple, respectively. At larger frequencies, the conductivity approaches a constant. The parameters were set to $\alpha=-1$, $\beta=0$, $\mu = 1.724$, $T=0.1$, $m=1$, $L=1$.}
\end{center}
\end{figure}

In the general case, the conductivity exhibits a Drude peak as seen in FIG. \ref{fig:pic1}.
The size of the peak grows as $m$ decreases. In the $m\to 0$ limit, we recover the delta function at $\om=0$.

\begin{figure*}
\begin{center}
\subfigure[\label{fig:picrnsigma}]{
\includegraphics[width=55mm]{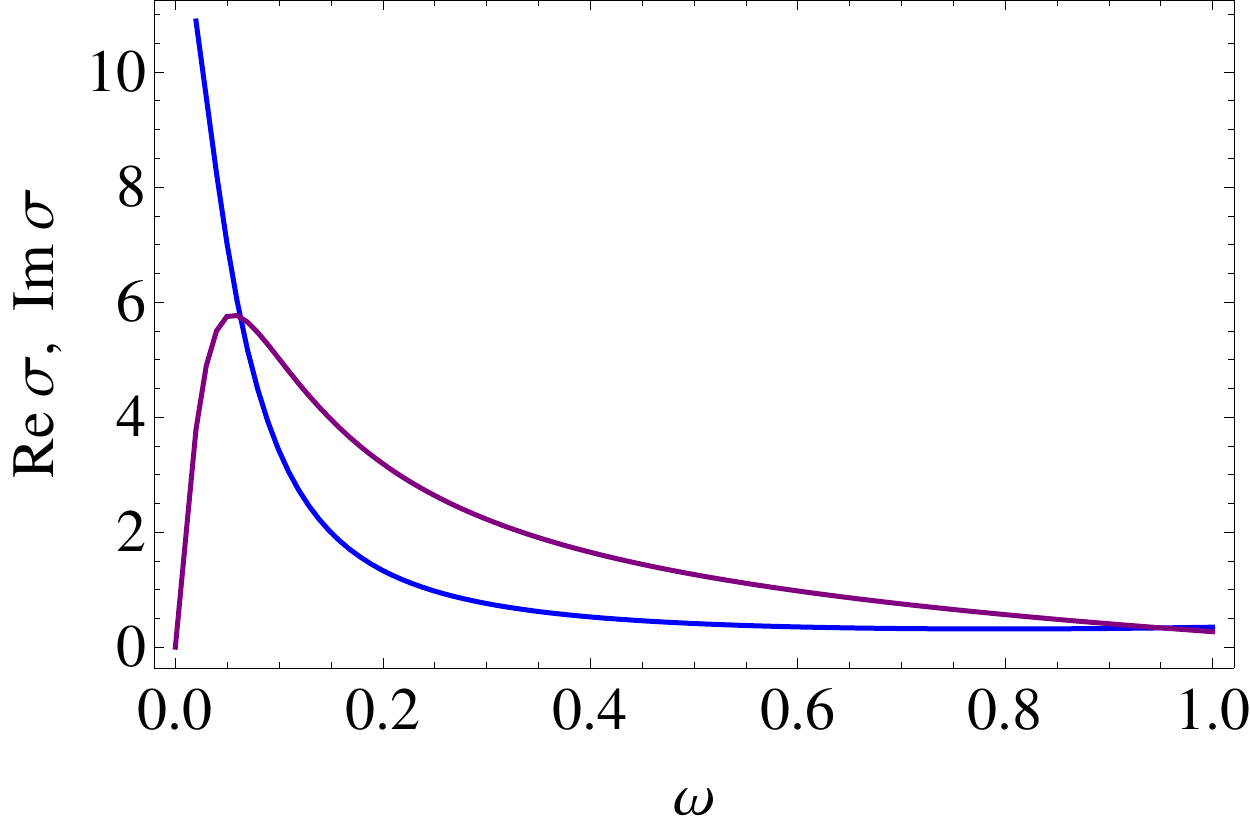}}
\subfigure[\label{fig:picrnexp}]{
\includegraphics[width=60mm]{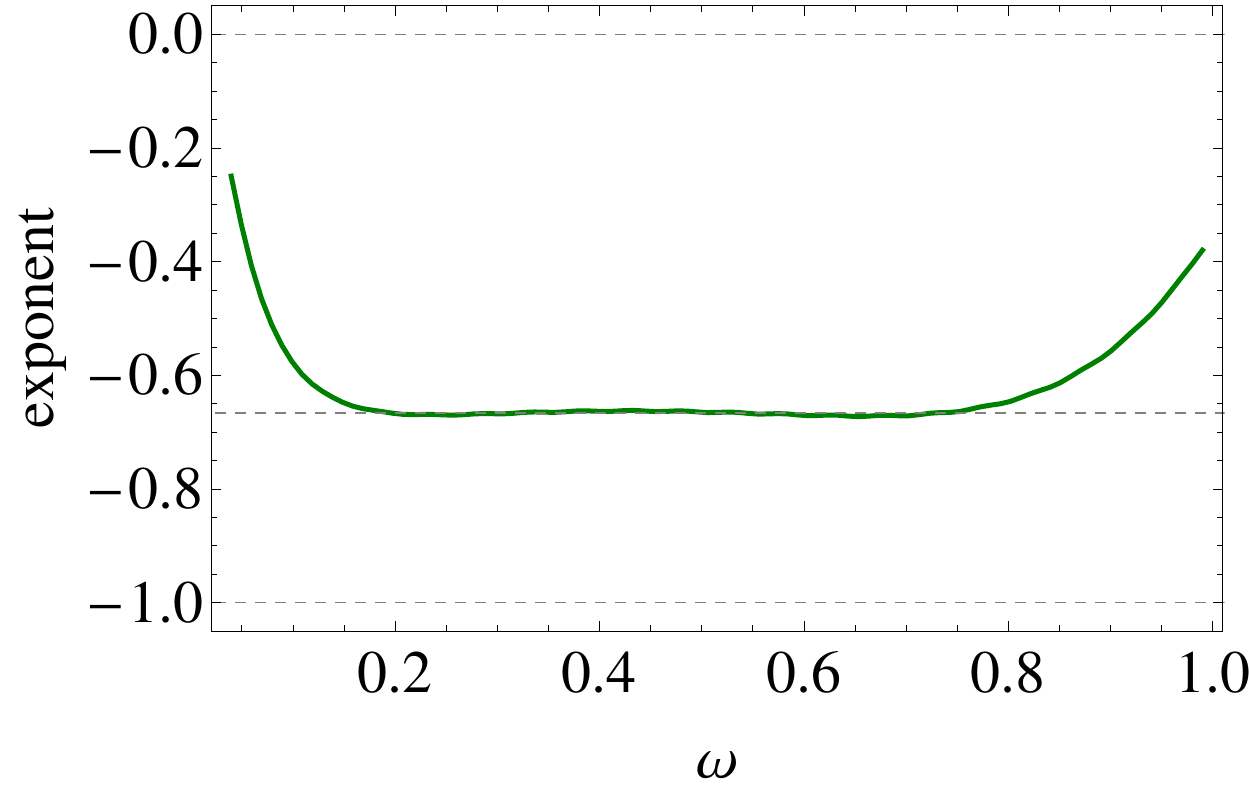}}
\subfigure[\label{fig:picrnphase}]{
\includegraphics[width=57mm]{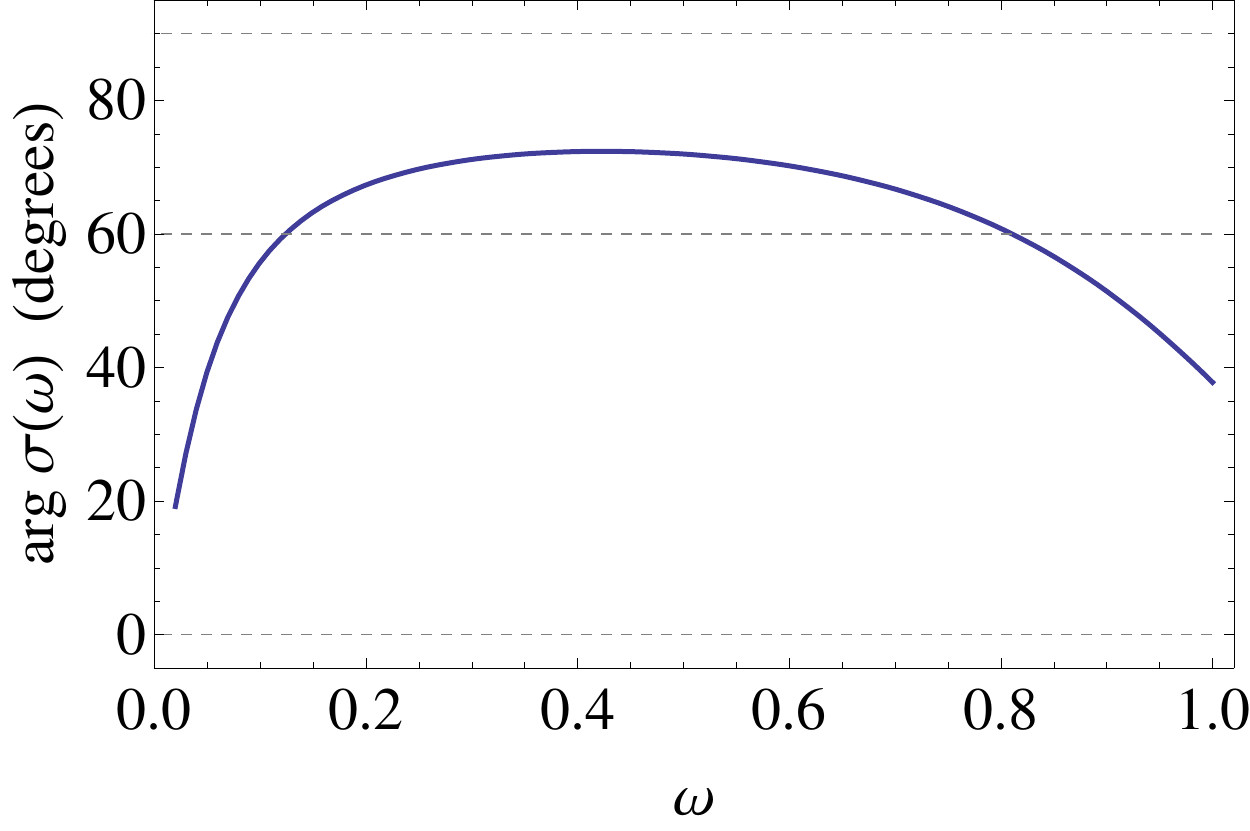}}
\caption{Non-Drude optical conductivity. There is an approximate power-law: $|\sigma(\om)| \approx {A \ov \om^{\gamma}} + B$. The mass is tuned to $L^2 m^2 \alpha = -0.75$ (and $\beta=0$) so that the exponent $\gamma\approx 2/3$ with an offset $B\approx -1.2$. The constants $\gamma$, $A$ and $B$ depend on the two parameters $\alpha$ and $\beta$. Fig.~\ref{fig:picrnsigma}: The blue and purple lines are the real and imaginary parts of the conductivity, respectively.  Fig.~\ref{fig:picrnexp}: The plot shows ${d\le(|\sigma|-B\ri) \ov d\le( \log \om \ri)}$ which gives the exponent if there is indeed a power law. Fig.~\ref{fig:picrnphase}: Phase of $\sigma(\om)$. If $B$ were zero, then it would exactly be $60^\circ$ due to causality and time-reversal invariance \cite{2003Natur.425..271M}.
}
\end{center}
\end{figure*}

\subsection{Stability}

In this paper we will not attempt to thoroughly study the conditions of stability.
There are certainly inconsistent regions in the parameter space, where the retarded gauge field correlator contains poles on the upper half-plane of complex frequencies. Numerical results indicate that this happens above the wall of stability  (see FIG. \ref{fig:psc}).
Between the wall of stability and the ``$\mu=0$'' line (the yellow region in FIG. \ref{fig:psc}) the system may be stable.

\subsection{Emergent non-Drude scaling}

The optical conductivity differs from the simple $|\sigma(\om)| \propto \om^{-1}$ that is predicted by the Drude theory. In an intermediate regime $T < \om < \mu$, we see an approximate behavior best described by
\be
   |\sigma(\om)| \approx {A \ov \om^{\gamma}} + B
\ee
where $\gamma$, $A$ and $B$ are $\mathcal{O}(1)$ constants that depend on the $\alpha$ and $\beta$ parameters. Numerical results gave $B<0$ in all cases. See FIGs. \ref{fig:picrnsigma}, \ref{fig:picrnexp}, \ref{fig:picrnphase} for a sample numerical solution. In these figures, we have tuned the graviton masses such that $\gamma \approx -2/3$.
The power law behavior extends to larger and larger regions as the temperature is lowered.
By changing the graviton masses, we obtain power laws with different exponents. As $m \to 0$, the peak becomes more and more Drude-like (i.e. $\gamma = 1$).
These results are very similar to those in \cite{Horowitz:2012ky, Horowitz:2012gs}.

\newpage

\section{Discussion}

In this paper, we studied massive gravity as a holographic framework for translational symmetry breaking and momentum dissipation.
Instead of directly dealing with inhomogeneous fields (e.g. the metric) and their perturbations, we considered averaged quantities that satisfy `renormalized' equations of motion. Ideally, figuring out what these modified equations are would be done by integrating out high-wavelength modes in the theory. However, that is a hard problem and instead we made a step by guessing their form by considering the symmetries of the system.

We arrived at a holographic theory of solids based on Lorentz-breaking graviton mass terms. We computed conductivities in different cases. The conductivity showed a Drude peak at zero frequency. We also observed non-Drude power-laws in the absolute value of the optical conductivity. These fat tails extended to frequencies comparable to the chemical potential. This ultraviolet effect seems to be unrelated to the physics that governs the DC conductivity, which in our model ultimately depended on the graviton mass only (at relatively small temperatures). It would be interesting to consider more general models which allow for a temperature-dependent DC conductivity.
In order to do this, the reference metric may be promoted to a dynamical quantity (see related work \cite{Hassan:2011zd}). For instance, if we consider the simple ansatz
\be
  f_{\mu\nu} = \textrm{diag}\le(0,0,F(r),F(r)\ri)
\ee
then the equation of motion for $F(r)$ gives
\be
  F(r) \propto g_{xx}(r) = {L^2\ov r^2}
\ee
This corresponds to a constant $\overline{m}^2(r)$ which is further equivalent to a shift in the cosmological constant.
One might also add a kinetic term and a mass term for $F(r)$ so that the $r$-dependence changes and it emulates  the finite-momentum `master field' in \cite{Hartnoll:2012rj}. It would be very important to develop a quantitative correspondence between lattice perturbations and graviton masses (and perhaps higher order corrections in the action).

We emphasize that we did not attempt to characterize the instabilities in these systems. There are certainly inconsistent regions in the parameter space, where the retarded gauge field correlator showed poles on the upper half-plane.  It would be important to understand under what conditions can the ghosts and tachyons be eliminated from the theory. Such investigations generally depend on the background.
Some instabilities are presumably associated with the growth of inhomogeneities with time (structure formation).

Instabilities generically lead to other phases.
A very interesting application of the results would be to study charged condensates as a model for supersolids or perhaps a pseudogap phase. It would  also  be interesting to find striped, dielectric, or insulating phases by changing the reference metric, or study electron stars \cite{Hartnoll:2010gu} in this context. One might also wonder whether there is a holographic analog of Cooper pairing and study how the effective phonon coupling changes in the radial direction.

One can try to extend the theory to other dimensions. In three spatial dimensions one might use homogeneous Bianchi metrics as a spatial reference metric which extends the number of possibilities. (For Bianchi spaces and holography, see e.g. \cite{Iizuka:2012iv, Donos:2012gg, Donos:2012wi, Iizuka:2012pn, Donos:2012js}.)

In the paper, general covariance was only broken in the spatial dimensions. It would be extremely valuable to develop a theory where time translations are also broken. This may be a first step toward constructing a holographic model of Kolmogorov's 1941 theory of fully developed turbulence.

Computations in holography might look complicated at first. The physics is often elucidated by a semi-holographic approach \cite{Faulkner:2010tq}. It would be very useful to develop an effective theory along the lines of \cite{Nickel:2010pr}.
At the technical level, analytical results would be extremely useful, perhaps by using
matched asymptotic expansions.  Finally, it would also be interesting to see if there are any relations to other, non-relativistic, holographic theories \cite{Kachru:2009xf}. In particular, how the conductivity calculations on Bianchi VII spaces that produce a Drude peak \cite{Donos:2012js} can be reformulated in our framework. This would presumably shed more light on the criteria of consistency in massive gravity theories.

As we have seen in section \ref{sec:bg}, thermodynamical state variables are modified by the finite graviton mass. In particular, entropy does not generically follow the usual $S=A/4$ law. It would be interesting to interpret these results and compute other related quantities (e.g. entanglement entropies).

In this paper, we substituted graviton mass terms for spatial inhomogeneities in asymptotically anti-de Sitter spacetime. As mentioned earlier in this section, graviton masses may cause a shift in the value of the effective cosmological constant that is seen by perturbations.
The relation of momentum dissipation and an effective cosmological constant can be demonstrated in a more direct way as follows. In the absence of external forces, we can write the algebra corresponding to the Drude model as
\be
 \nonumber
  {d \over dt}  P_i  = \{ P_i, H \} = -{P_i \ov \tau} \qquad  \quad \{ P_i, P_j \} = 0
\ee
where $H$ is the Hamiltonian and $\{\cdot,\cdot\}$ is the Poisson bracket.
In 2+1 dimensions, the algebra spanned by $\{H, P_x, P_y\}$ corresponds to Bianchi type V spaces where time plays the role of one of the three Bianchi dimensions.
The simplest example for a spacetime whose Killing vectors obey this algebra is de Sitter space
\be
  ds^2 = -dt^2 + e^{-t/\tau} d\vec x^2
\ee
and a corresponding cosmological constant is given by $\Lambda = 3/(2\tau^2)$.
We see two dual pictures emerging. We either have flat space with momentum dissipation, or de Sitter with conserved momentum. In the latter, we have traded momentum dissipation for the expansion of space: momentum is simply being inflated away.

Using these ideas, one can calculate a `mean free path' corresponding to the cosmological constant of our Universe. We get $\lambda = c\tau \approx 3.4$ Gpc. Amusingly, this is only a few dozen times bigger than the 100 Mpc `lattice spacing' of the large-scale structures (above which the Universe is approximately homogeneous and isotropic).

\vspace{0.2in}   \centerline{\bf{Acknowledgments}} \vspace{0.2in} I benefited from discussions with Lasma Alberte, Diego Blas, Aristomenis Donos, Jerome Gauntlett, Sean Hartnoll,  Diego Hofman, Gary Horowitz, Shamit Kachru, Subodh Patil, Massimo Porrati, Jorge Santos, and David Tong. I would like to thank Shamit Kachru, John McGreevy, and Sean Hartnoll for helpful comments on the manuscript.
I further thank Harvard University, the Newton Institute, Imperial College, Stanford University and SLAC for hospitality.

\appendix

\bwt

\section{Equations for the background}

Einstein's equations are supplemented by the graviton mass term,
\be
  R_{\mu\nu} - {R\ov 2} g_{\mu\nu} + \Lambda  g_{\mu\nu} + F_{\mu\alpha}F^{\alpha}_\nu + { g_{\mu\nu} \ov 4} F_{\alpha\beta}F^{\alpha\beta} + m^2 X_{\mu\nu}=0
\ee
where $\Lambda = 6/L^2$ and
\be
   X_{\mu\nu}= {\alpha \ov 2} \le(  [\mathcal{K}] g_{\mu\nu} -  \mathcal{K}_{\mu\nu} \ri) -
   {\beta} \le(  (\mathcal{K}^2)_{\mu\nu}  - [\mathcal{K}]  \mathcal{K}_{\mu\nu} + \half  g_{\mu\nu}\le( [\mathcal{K}]^2 - [\mathcal{K}^2] \ri) \ri)
\ee
with ${\mathcal{K}^\mu}_\nu (g, f) = {(\sqrt{g^{-1}f})^\mu}_\nu$. Indices are lowered and raised by the metric $g$. The $t-t$ component gives a differential equation for the emblackening factor
\be
2 r f'(r)-6 f(r)+2 \alpha  F L m^2 r+2 \beta  F^2 m^2 r^2-\frac{\mu ^2 r^4}{2 r_h^2}+6 =0
\ee
whose solution for $f(r)$ is the one given in Section \ref{sec:bg}.

\section{Equations for perturbations}

At the linear level, we obtain the following three equations
{
\be
  \nonumber
2 L^2 r_h \omega ^2 a_x(r)+f(r) \left(2 L^2 r_h f(r) a_x''(r)+2 L^2 r_h a_x'(r) f'(r)+i \mu  r^2 \omega  g_{rx}(r)-\mu  r^2 g_{tx}'(r)-2 \mu  r g_{tx}(r)\right) =0
\ee
\bea
  \nonumber
  & g_{tx}(r) \left(-2 r^2 r_h^2 f''(r)+8 r r_h^2 f'(r)-16 r_h^2 f(r)+4 \alpha  F L m^2 r r_h^2+4 \beta  F^2 m^2 r^2 r_h^2+\mu ^2 r^4+12 r_h^2\right)+ & \\
  \nonumber
  & +2 r r_h f(r) \left(-2 \mu  L^2 r a_x'(r)+r_h \left(-i r \omega  g_{rx}'(r)+r g_{tx}''(r)+2 g_{tx}'(r)\right)\right)=0 &
\eea
\bea
  \nonumber
  &  g_{rx}(r) \left(f(r) \left(-2 r_h^2 \left(r^2 f''(r)-4 r f'(r)+6 f(r)\right)+4 r_h^2 \left(\alpha  F L m^2 r+\beta  F^2 m^2 r^2+3\right)+\mu ^2 r^4\right)+2 r^2 r_h^2 \omega ^2\right)+ & \\
  \nonumber
  & +2 i r r_h^2 \omega  \left(r g_{tx}'(r)+2 g_{tx}(r)\right) -4 i \mu  L^2 r^2 r_h \omega  a_x(r) =0 &
\eea
}

Note that the equations become real if we multiply $g_{rx}$ by $i$. If $m=0$, then we can consistently set $g_{rx}=0$ and then the second equation becomes dependent on the third one. Since $m>0$ only introduces $g_{tx}$, but not its derivatives, we can express $g_{tx}$ using the other variables.

\ewt

\bibliography{drude}

\end{document}